**Spin reorientation in the square-lattice antiferromagnets RMnAsO (R = Ce, Nd): Density functional analysis of the spin exchange interactions between the rare-earth and transition-metal ions**

Changhoon Lee[1], Erjun Kan[2], Hongjun Xiang[3], Reinhard K. Kremer,[4] Seung-Hun Lee[5], Zenji Hiroi[6] and Myung-Hwan Whangbo[1,*]

[1] Department of Chemistry, North Carolina State University, Raleigh, NC 27695-8204, USA

[2] Department of Applied Physics, Nanjing University of Science and Technology, Nanjing, Jiangsu 210094, P. R. China

[3] Key Laboratory of Computational Physical Sciences (Ministry of Education), and Department of Physics, Fudan University, Shanghai 200433, P. R. China

[4] Max-Planck-Institut für Festkörperforschung, Heisenbergstr. 1, D-70569 Stuttgart, Germany

[5] Department of Physics, University of Virginia, Charlottesville, Virginia 22904-4714, USA

[6] Institute for Solid State Physics, University of Tokyo, Kashiwa, Chiba 277-8581, Japan

**Abstract**

The spin reorientation (SR) phenomenon of the square-lattice antiferromagnets RMnAsO (R = Ce, Nd) was investigated by analyzing the spin exchange interactions between the rare-earth and transition-metal ions ($R^{3+}$ and $Mn^{2+}$, respectively) on the basis of density functional calculations. It is found that the symmetry and strength of the Dzyaloshinskii-Moriya (DM) interaction are determined primarily by the partially filled 4f states of the $R^{3+}$ ions, and that the DM and biquadratic (BQ) exchanges between the $R^{3+}$ and $Mn^{2+}$ ions are unusually strong and control the observed spin reorientation phenomenon. Below their SR temperature, the $Mn^{2+}$ and $Ce^{3+}$ moments are orthogonal in CeMnAsO but are collinear in NdMnAsO, because the DM interaction dominates over the BQ interaction for CeMnAsO while the opposite is the case for NdMnAsO. Experiments designed to test the implications of our findings are proposed.

**1. Introduction**

The energy spectrum of a magnetic solid consisting of ions with unpaired spins is commonly described by considering the isotropic bilinear (Heisenberg), the antisymmetric (Dzyaloshinskii-Moriya), and anisotropic exchange between various pairs of its spin sites,[1,2] although some inadequacy of this approach has been pointed out in the studies of $La_2CuO_4$.[3] For magnetic ions with spin S > 1/2, it may become necessary to rectify the inadequacy of using the Heisenberg exchange $\sum_{i<j} J_{ij} \vec{S}_i \cdot \vec{S}_j$ by including the biquadratic (BQ) exchange $\sum_{i<j} K_{ij} (\vec{S}_i \cdot \vec{S}_j)^2$.[4-6] The Dzyaloshinskii-Moriya (DM) exchange $\sum_{i<j} \vec{D}_{ij} \cdot (\vec{S}_i \times \vec{S}_j)$ is responsible for a variety of interesting magnetic phenomena such as spin canting in antiferromagnets,[2,7] ferroelectric polarization induced by chiral magnetic order [8,9] and magnon quantum Hall effect,[10] to name a few. In general, the strength of the DM exchange is often estimated by $|D_{ij}/J_{ij}| \approx \Delta g/g$, where $\Delta g$ is the contribution of the orbital moment to the g-factor g,[11,12] and is commonly believed to be at most 1/10 of the Heisenberg exchange. Similarly, the strength of the anisotropic exchange is estimated by $(\Delta g/g)^2$ and is at most 1/10 of the DM exchange.[11,12]

Exchange interactions between TM and rare-earth (RE) ions have been less extensively studied than those between TM ions.[13] So far, studies on the DM exchange have been mostly concerned with interactions among identical TM ions. Consequently, the DM exchange between TM and RE ions is not well understood. The anisotropic exchange [1,11] between RE and TM ions has been examined in connection with the spin reorientation (SR) phenomenon, for example, involving the $Nd^{3+}(f^3)$ and $Cu^{2+}(d^9)$ moments in the square lattice antiferromagnet $Nd_2CuO_4$,[14,15] in which two $Nd^{3+}$ ions are located above and below every square of $Cu^{2+}$ ions forming a $Cu_4Nd_2$ square bipyramid (**Fig. 1a**). In the ordered magnetic states of $Nd_2CuO_4$ (**Fig. 1b**),[14] in which the $Cu^{2+}$ ions form a square antiferromagnetic lattice, the net Heisenberg exchange between the $Nd^{3+}$ and $Cu^{2+}$ ions vanishes. The DM exchange between them also vanishes because the DM interaction of one $Nd^{3+}$ ion lying above each $Cu_4$ square is canceled by that of the other $Nd^{3+}$ ion located below the $Cu_4$ square. A nonzero DM exchange would have resulted if the two $Nd^{3+}$ ions lying above and below each $Cu_4$ square had opposite spin directions. Thus, the aniostropic exchange[1,11] is left as the only remaining exchange, provided that the BQ exchange is negligible.

A nonzero DM interaction between RE and TM ions can be realized in the square-lattice antiferromagnets NdMnAsO [16,17] and PrMnSbO,[18] which are analogous in their crystal structure to the iron arsenide superconductors.[19,20] In RMnAsO (R = Nd, Pr), each square sheet of high-spin $Mn^{2+}$ ($d^5$, S = 5/2) ions (parallel to the ab-plane) is sandwiched between two square sheets of $As^{3-}$ ions to form an As-Mn-As slab (**Fig. 2a**), and each square sheet of $O^{2-}$ ions (parallel to the ab-plane) between two square sheets of $R^{3+}$ ions to form a R-O-R slab (**Fig. 2a**). The As-Mn-As slabs alternate with the R-O-R slabs along the c-direction such that each R is located above the center of an $As_4$ square, and each As above the center of a $R_4$ square (**Fig. 2a**). Consequently, the RE ions $R^{3+}$ ions lying above and below the square sheet of the $Mn^{2+}$ ions occupy the positions of neighboring $Mn_4$ squares, with each R atom forming $Mn_4R$ square pyramid (**Fig. 2b**). The $Mn^{2+}(d^5)$ ions in the square lattice of RMnAsO (R = Nd, Pr) undergo an antiferromagnetic ordering at high temperatures ($T_N$ = 359 and 230 K for R = Nd and Pr, respectively) with the Mn moments oriented along the c-axis.[16] NdMnAsO and PrMnSbO undergo a SR transition at low temperatures ($T_{SR}$ = 23 K and 35 K, respectively), below which the Mn moments rotate into the ab-plane. Simultaneously, the RE moments occur in the ab-plane with ferromagnetic coupling in each sheet of RE atoms parallel to the ab-plane, and two such sheets straddling each sheet of the Mn atoms have their moments antiferromagnetically coupled (**Fig. 3**). CeMnAsO consists of $Ce^{3+}(f^1)$ and $Mn^{2+}(d^5)$ ions, is isostructural with NdMnAsO, and undergoes an analogous SR at $T_{SR}$ = 34 K.[21,22] LaMnAsO exhibits an antiferromagnetic ordering of the Mn moments similar, as found RMnAsO (R = Ce, Nd). However, with no moment on $La^{3+}$, LaMnAsO does not undergo a SR on cooling and keeps the Mn moments aligned along the c axis.[17] This indicates that the SR in NdMnAsO and CeMnAsO is caused by the interaction between the $R^{3+}$ and $Mn^{2+}$ ions. There is an important difference between the magnetic structures of CeMnAsO and NdMnAsO below $T_{SR}$, namely, the $R^{3+}$ and $Mn^{2+}$ moments are orthogonal in CeMnAsO [21,22] but are collinear in NdMnAsO.[17]

In the present work we probe the cause for the SR of RMnAsO (R = Ce, Nd) on the basis of density functional theory (DFT) calculations. We show that the DM and BQ exchange interactions between the $R^{3+}$ and $Mn^{2+}$ ions are strong and govern the nature of their SR phenomenon, and that the DM interaction dominates over the BQ interaction in CeMnAsO, but the opposite is the case for NdMnAsO. In addition, we explore the implications of our findings that can be tested by further experiments and theoretical analyses.

## 2. Interactions between rare-earth and transition-metal ions

### A. Nature of the DM exchange

Let us examine under what condition the DM exchange between the $R^{3+}$ and $Mn^{2+}$ ions of RMnAsO (located respectively at sites 0 and i of the $Mn_4R$ square pyramid, **Figs. 3a** and **3c**) can become strong. The DM exchange between $R^{3+}$ and $Mn^{2+}$ in a $Mn_4R$ square pyramid is written as $\vec{D}_{0i} \cdot (\vec{S}_0 \times \vec{S}_i)$ (i = 1 – 4) with the DM vector expressed as $\vec{D}_{0i} = \lambda J_{0i}(\delta\vec{L}_0 - \delta\vec{L}_i)$,[2,23] where $\lambda$ is the spin-orbit coupling (SOC) constant of the magnetic ions, while $\delta\vec{L}_0$ and $\delta\vec{L}_i$ are essentially the unquenched orbital momenta of the magnetic ions at 0 and i, respectively.[23] Thus, the magnitude of $\vec{D}_{0i}$ can be increased by increasing the $(\delta\vec{L}_0 - \delta\vec{L}_i)$ value. For a S=5/2 TM ion (e.g., $Mn^{2+}$), L = 0 to a first approximation so that $\delta\vec{L} \approx 0$. For a RE ion $R^{3+}$, the quenching of the orbital angular momentum is generally weak. For the f-orbitals of a $R^{3+}$ ion overlap poorly with the orbitals of its surrounding ligands so that its spherical electron density distribution is weakly distorted by the surrounding ligands. Therefore, the unquenched orbital momentum $\delta\vec{L}$ of a RE ion $R^{3+}$ can be large. Therefore, the DM vector for the exchange between such TM and RE ions can be approximated by

$$\vec{D}_{0i} \approx \sqrt{\lambda_0 \lambda_i}\ J_{0i}(\delta\vec{L}_0 - \delta\vec{L}_i) \approx \sqrt{\lambda_0 \lambda_i}\ J_{0i}\delta\vec{L}_0. \tag{1}$$

Since the SOC constant of 4f ions is about three times larger than that of 3d ions,[24] such combinations of RE and TM ions would increase the length of the associated DM vector. The use of a S=5/2 TM ion is also favorable in strengthening the DM exchange, because it can increase the length of the vector $(\vec{S}_0 \times \vec{S}_i)$.

When the spins of the RE and TM ions are both large, the BQ exchange can become strongly enhanced because it is proportional to $(\vec{S}_0 \cdot \vec{S}_i)^2$. Therefore, it is an important issue whether the DM or BQ exchange dominates, because the DM exchange favors an orthogonal arrangement between the RE and TM moments whereas the BQ exchange favors a collinear arrangement.

**B. DM and BQ exchange energies**

In their powder neutron diffraction study of NdMnAsO,[16] Marcinkova *et al*. were unable to resolve whether the ordered Nd and Mn moments below $T_{SR}$ = 23 K, which lie in the ab-plane, are collinear (**Fig. 3b**) or orthogonal (**Fig. 3c** or **3d**). However, Emery *et al*.[17] have recently reported that the ordered Mn and Nd moments below $T_{SR}$ have the collinear arrangement depicted in **Fig. 3b**. For the convenience of our discussion, the ab-plane collinear arrangement in **Fig. 3b** will be referred to as the //ab-C arrangement, the ab-plane orthogonal arrangement in **Fig. 3c** as the //ab-O1 arrangement, and the alternative ab-plane orthogonal arrangement in **Fig. 3d** as the //ab-O2 arrangement. The latter arrangement results from the //ab-O1 arrangement by reversing the directions of the RE moments while keeping the Mn moments unchanged. The recent neutron powder diffraction study [17] showed that the Mn moments of LaMnAsO at room temperature are parallel to the c-axis, as found for those of NdMnAsO between $T_{SR}$ and $T_N$.

We now consider the energies of the ab//-C, ab//-O1 and ab//-O2 spin arrangements of RMnAsO (R = Ce, Nd), which we denote as E(ab//-C), E(ab//-O1) and E(ab//-O2), respectively, in terms of the spin Hamiltonian defined by the Heisenberg, BQ and DM spin exchange interactions

$$H_{spin} = \sum_{i<j} J_{ij} \vec{S}_i \cdot \vec{S}_j + \sum_{i<j} K_{ij} (\vec{S}_i \cdot \vec{S}_j)^2 + \sum_{i<j} \vec{D}_{ij} \cdot (\vec{S}_i \times \vec{S}_j). \qquad (2)$$

(With the above definition, ferromagnetic and antiferromagnetic Heisenberg exchanges are given by $J_{ij} < 0$ and $J_{ij} > 0$, respectively.) Then, the net Heisenberg exchange between the $R^{3+}$ and $Mn^{2+}$ ions is zero for the //ab-C arrangement, and also for the ab//-O1 and ab//-O2 arrangements due to their orthogonal spin arrangement. For the ab//-C arrangement (**Fig. 3b**), the DM exchange energy ($E_{DM}$) is zero but the BQ exchange energy ($E_{BQ}$) is not, because the $R^{3+}$ and $Mn^{2+}$ moments are collinear. For the //ab-O1 and //ab-O2 arrangements, $E_{BQ} = 0$ because the $R^{3+}$ and $Mn^{2+}$ ions moments are orthogonal, and their $E_{DM}$ energies are nonzero and are opposite in sign. Therefore,

$$2|E_{DM}| = |E(//ab\text{-}O1) - E(//ab\text{-}O2)| \qquad (3)$$

Let us use the notation //ab-$O_l$ to indicate the lower-energy one of the //ab-O1 and //ab-O2 arrangements. Then the energy difference between the //ab-C and //ab-$O_l$ states has both the BQ and the DM contributions. Thus,

$$E_{BQ} = E(//ab\text{-}C) - E(//ab\text{-}Ol) - |E_{DM}| \qquad (4)$$

Consequently, one can evaluate $E_{DM}$ and $E_{BQ}$ by determining the relative energies of the //ab-C, //ab-O1 and //ab-O2 arrangements on the basis of DFT+U+SOC calculations.

## 3. Analysis of the electronic and magnetic structures

### A. Computational details

We examine the electronic structures of LaMnAsO, CeMnAsO and NdMnAsO on the basis of DFT electronic structure calculations. For LaMnAsO, we employed the frozen-core projector augmented wave method encoded in the Vienna Ab initio Simulation Package [25] with the generalized-gradient approximation [26] for the exchange-correlation functional, the plane-wave cut-off energy of 400 eV, and 27 k-points for the irreducible Brillouin zone. To describe the electron correlation in the Mn 3d states, the DFT plus on-site repulsion U (DFT+U) method [27] was used with effective on-site Coulomb repulsion U = 4.5 eV on the Mn atom, a representative value for Mn.[28,29] The threshold for the self-consistent-field convergence of the total electronic energy was $10^{-6}$ eV.

To evaluate the Heisenberg exchange interactions between the $R^{3+}$ and $Mn^{2+}$ ions in RMnAsO (R = Ce, Nd), it is necessary to treat the f-electron of the $R^{3+}$ ions explicitly. Thus, for RMnAsO (R = Ce, Nd), we carried out DFT+U calculations (with U = 4.5 and 5.4 eV on the Mn and R atoms, respectively[30]) by using the full-potential linearized augmented plane wave method encoded in the WIEN2k package [31] with 64 k-points for the irreducible Brillouin zone, the threshold of $10^{-6}$ Ry for the energy convergence, the cut-off energy parameters of $RK_{max} = 7.0$ and $G_{max} = 12$, and the energy threshold of -6.0 Ry for the separation of the core and valence states. We use the following sets of basis orbitals: [Ar] $3d^5 4s^2$ for Mn, [Xe] $4f^2 6s^2$ for Ce, [Ar] $3d^{10}4s^2 4p^3$ for As, and [He] $2s^2 2p^4$ for O. The muffin-tin sphere radii used are 2.5, 2.34, 2.25, and 2.08 for Mn, Ce, As, and O, respectively. To determine the spin orientations of the $R^{3+}$ and $Mn^{2+}$ ions and evaluate the DM interaction energies of RMnAsO (R = Ce, Nd), we perform DFT+U plus SOC [32] (DFT+U+SOC) calculations.

## B. Electronic structures

The total density of states (DOS) plots obtained from the DFT+U+SOC calculations for the //ab-C spin arrangement of CeMnAsO and NdMnAsO are presented in **Figs. 4a** and **5a**, respectively. The partial DOS (PDOS) plots obtained for the Mn 3d and the Ce 4f states are given **Figs. 4b** and **4c**, respectively, and those obtained for the Mn 3d and the Nd 4f states in **Figs. 5b** and **5c**, respectively. The upper and lower panels in each diagram refer to the up-spin and down-spin states, respectively. Both CeMnAsO and NdMnAsO exhibit a band gap, and hence are magnetic insulators. The PDOS plots for the Mn 3d states show that the up-spin states are all filled but the down-spin states are all empty, consistent with the presence of high-spin $Mn^{2+}$ ions (i.e., S = 5/2) in RMnAsO (R = Ce, Nd). The PDOS plots for the Ce 4f and Nd 4f states are consistent with the $Ce^{3+}(f^1)$ and $Nd^{3+}(f^3)$ electron counts, respectively (for further discussion, see below).

## C. Magnetic anisotropy of individual magnetic ions

On the basis of DFT+U+SOC calculations we first estimate the magnetic anisotropy of the individual $Mn^{2+}$, $Ce^{3+}$ and $Nd^{3+}$ ions in RMnAsO (R = Ce, Nd) in the absence of interactions between $R^{3+}$ and $Mn^{2+}$ ions. Our DFT+U+SOC calculations for the magnetic ground state of LaMnAsO, in which the Mn moments in the square lattices are antiferromagnetically ordered, show that the $Mn^{2+}$ moment parallel to the c-axis (//c) is more stable than that perpendicular to the c-axis ($\perp$c) by 0.20 meV per Mn, which is consistent with the experimental observations for LaMnAsO at room temperature and for NdMnAsO and PrMnSbO between $T_{SR}$ and $T_N$.[16-18] Our DFT+U+SOC calculations for the structure RZnAsO, resulting from RMnAsO when the magnetic $Mn^{2+}$ ions are replaced with the nonmagnetic $Zn^{2+}$ ions, show that the $Ce^{3+}$ spin moment favors the //c orientation over the $\perp$c orientation by 3.5 meV per Ce. In contrast, the $Nd^{3+}$ spin moment prefers the $\perp$c orientation to the //c orientation by 20.7 meV per Nd. Thus, the $Mn^{2+}$ ions of RMnAsO have weak easy-axis anisotropy as may be expected for the L=0 ion $Mn^{2+}$. The $Ce^{3+}$ ions of CeMnAsO have substantial easy-axis anisotropy. In contrast, the $Nd^{3+}$ ions of NdMnAsO exhibit strong easy-plane anisotropy. For CeMnAsO to undergo a SR with a rotation of the Mn moments into the ab-plane, the interaction between the $Ce^{3+}$ and $Mn^{2+}$ ions needs to overcome the easy-axis anisotropy of the individual Ce and Mn moments.

The single-ion magnetic anisotropy of a TM ion is a weak effect arising from SOC. As found for numerous magnetic systems such as $Ag_2MnO_2$,[33] $TbMnO_3$,[34] $SrFeO_2$,[35] $Sr_3Fe_2O_5$,[36] not to mention LaMnAsO examined in the present work, this anisotropy is well reproduced by DFT+U+SOC calculations despite the inherent limitations of the exchange-correlation functionals. This indicates that the deficiency of the functional cancel out when the energy difference between different spin orientations is calculated.

## D. Energy mapping analysis for the DM and BQ exchange energies

The relative energies calculated for the //a-C, //ab-O1 and //ab-O2 arrangements of RMnAsO (R = Ce, Nd) are summarized in **Table 1**, and so are the values of $E_{DM}$ and $E_{BQ}$ of RMnAsO (R = Ce, Nd) calculated from the relative energies by using Eqs. 3 and 4. For CeMnAsO, the orthogonal arrangement //ab-O1 is the ground state, and the DM exchange dominates over the BQ exchange, with $E_{DM} = -4.88$ meV per formula unit (FU) and $E_{BQ} = -0.07$ meV/FU. The DM exchange favoring the //ab-O1 arrangement overrides not only the BQ exchange favoring the //ab-C arrangement but also the easy-axis anisotropy of the individual $Ce^{3+}$ and $Mn^{2+}$ ions (i.e., 3.7 meV/FU). Therefore, the SR of CeMnAsO is caused by the DM exchange between the $Ce^{3+}$ and $Mn^{2+}$ ions. For NdMnAsO, the collinear arrangement //ab-C is the ground state, and the BQ exchange dominates over the DM exchange, with $E_{BQ} = -18.05$ meV/FU and $E_{DM} = -0.35$ meV/FU. Thus, the SR of NdMnAsO is induced by the strong easy-plane anisotropy of the $Nd^{3+}$ ion plus the strong BQ exchange between the $Nd^{3+}$ and $Mn^{2+}$ ions.

## E. Heisenberg spin exchange between the $R^{3+}$ and $Mn^{2+}$ ions

For our discussion in the next section, it is necessary to evaluate the Heisenberg spin exchange constant $J_{0i}$ between the $R^{3+}$ and $Mn^{2+}$ ions (Eq. 1). As depicted in **Fig. 6**, one may consider the five Heisenberg exchange paths $J_1 - J_5$ in RMnAsO (R = Ce, Nd), where $J_{0i}$ is defined as $J_5$. To extract the values of $J_1 - J_5$, we determine the relative energies of the six ordered spin states FM and AF1 – AF5 depicted in **Fig. 7** by DFT+U calculations. Given the spin Hamiltonian

$$\hat{H} = \sum_{i<j} J_{ij} \hat{S}_i \cdot \hat{S}_j \tag{5}$$

defined in terms of the spin exchanges $J_{ij} = J_1 - J_5$, the total spin exchange interaction energies of the FM and AF1 – AF5 states (per two FUs) are obtained as summarized below by applying the energy expressions obtained for spin dimers [37]

$$E_{FM} = (+4J_1 + 4J_2)N^2/4 + (+4J_3 + 4J_4)M^2/4 + (+8J_5)NM/4$$

$$E_{AF1} = (-4J_1 + 4J_2)N^2/4 + (+4J_3 + 4J_4)M^2/4$$

$$E_{AF2} = (-4J_2)N^2/4 + (+4J_3 + 4J_4)M^2/4 \qquad (6)$$

$$E_{AF3} = (+4J_1 + 4J_2)N^2/4 + (+4J_3 + 4J_4)M^2/4 + (-8J_5)NM/4$$

$$E_{AF4} = (-4J_1 + 4J_2)N^2/4 + (-4J_3 + 4J_4)M^2/4$$

$$E_{AF5} = (-4J_1 + 4J_2)N^2/4 + (-4J_4)M^2/4$$

where N and M are the numbers of the unpaired spins at the $Mn^{2+}$ and $R^{3+}$ sites of RMnAsO (R = Ce, Nd). Namely, N = 5 and M = 1 for CeMnAsO, and N = 5 and M = 3 for NdMnAsO. Thus, by mapping the relative energies of the six ordered spin states determined from the DFT+U calculations onto the corresponding energies obtained from the total Heisenberg exchange energies,[38] we obtain the $J_1 - J_5$ values of CeMnAsO summarized in **Table 2** indicating that the Heisenberg exchange between the $R^{3+}$ and $Mn^{2+}$ ions in CeMnAsO is ferromagnetic (−11.2 meV). According to Eq. 5, $J_5$ can be determined by considering only the FM and AF3 states. This is how we determine $J_5$ is antiferromagnetic (+38.4 meV) for NdMnAsO. Therefore, $J_5$ is larger in magnitude for NdMnAsO than for CeMnAsO. We note that DFT+U calculations for magnetic compounds made up of TM ions are known to overestimate Heisenberg spin exchange constants typically by a factor up to four to five depending upon what U values are used, but the trends in their relative values remain nearly independent of the U values used in the DFT+U

calculations.[39-42]

## 4. Discussion

### A. Nature of the DM exchange energy

To see why the DM exchange is strong in CeMnAsO but weak in NdMnAsO, we examine the nature of the DM exchange on the basis of Eq. 1. Consider the DM interaction around one $R^{3+}$ ion of RMnAsO in the //ab-O1 arrangement (**Fig. 3c**), where the site 0 is used for the $R^{3+}$ ion and the sites 1 – 4 for the $Mn^{2+}$ ions. Then the $E_{DM}$ is expressed as

$$E_{DM} = \vec{D}_{01} \cdot (\vec{S}_0 \times \vec{S}_1) + \vec{D}_{02} \cdot (\vec{S}_0 \times \vec{S}_2) + \vec{D}_{03} \cdot (\vec{S}_0 \times \vec{S}_3) + \vec{D}_{04} \cdot (\vec{S}_0 \times \vec{S}_4)$$
$$= (\vec{D}_{01} - \vec{D}_{02} + \vec{D}_{03} - \vec{D}_{04}) \cdot (\vec{S}_0 \times \vec{S}_1) \qquad (8)$$

Eq. 1 shows that the symmetry of the DM vectors $\vec{D}_{0i}$ (i = 1 – 4) is determined largely by that of the unquenched orbital momentum $\delta\vec{L}_0$ of the $R^{3+}$ ion, namely, by the symmetry of the occupied 4f orbitals of the $R^{3+}$ ion. Our DFT+U+SOC calculations for CeMnAsO show that the occupied 4f-states of the $Ce^{3+}$ ion have only the $l_z = \pm 2$ orbitals (**Fig. 8a**). In Cartesian coordinates, the latter correspond to the $f_{xyz}$ and $f_{z(x2-y2)}$ orbitals, which switch their sign when rotated around the z-axis by 90°. This means that $\vec{D}_{01} = \vec{D}_{03} = -\vec{D}_{02} = -\vec{D}_{04}$, so that

$$E_{DM} = 4\vec{D}_{01} \cdot (\vec{S}_0 \times \vec{S}_1). \qquad (9)$$

Thus, for the //ab-O2 arrangement (**Fig. 3d**), $E_{DM} = -4\vec{D}_{01} \cdot (\vec{S}_0 \times \vec{S}_1)$.

Our calculations for NdMnAsO show that the occupied 4f-states of the $Nd^{3+}$ ion have a strong contribution from the $l_z = \pm 1$ orbitals, but weak contributions from the $l_z = \pm 2$ and $l_z = \pm 3$ orbitals (**Fig. 8b**). The $l_z = \pm 1$ orbitals, corresponding to the $f_{x(4z2-x2-y2)}$ and $f_{y(4z2-x2-y2)}$ orbitals in Cartesian coordinates, switch their sign when rotated around the z-axis by 180° so that $\vec{D}_{01} =$ $-\vec{D}_{03}$ and $\vec{D}_{02} = -\vec{D}_{04}$. Consequently, the DM exchange associated with these orbitals is zero.

The $l_z = \pm 2$ orbitals will give rise to $E_{DM} = 3n_{\pm 2}(-4\vec{D}_{01} \cdot \vec{S}_c)$, where $3n_{\pm 2}$ represents the number of 4f electrons in the $l_z = \pm 2$ orbitals. The $l_z = \pm 3$ orbitals, which correspond to the $f_{x(x2-3y2)}$ and $f_{y(3x2-y2)}$ orbitals, switch their sign when rotated around the z-axis by 120°. If it is assumed that the DM vectors associated with the $l_z = \pm 3$ orbitals change linearly as a function of the rotation angle around the z-axis, one might expect that $\vec{D}_{02} = -\vec{D}_{01}/2$, $\vec{D}_{03} = 0$, and $\vec{D}_{04} = \vec{D}_{01}/2$. Thus, the DM exchange energy associated with the $l_z = \pm 3$ orbitals would be $3n_{\pm 3}(-\vec{D}_{01} \cdot \vec{S}_c)$, where $3n_{\pm 3}$ represents the number of 4f electrons in the $l_z = \pm 3$ orbitals. The DM exchange of NdMnAsO is small due to the low occupations of the $l_z = \pm 2$ and $l_z = \pm 3$ orbitals. This explains why the DM exchange is weak in NdMnAsO.

Eq. 1 shows that the sign of the DM vector $\vec{D}_{0i}$ between the $R^{3+}$ and $Mn^{2+}$ ions depends on the sign of their Heisenberg exchange constant $J_{0i}$. As noted above, $J_{0i}$ is ferromagnetic ($J_{0i} < 0$) in CeMnAsO but antiferromagnetic ($J_{0i} > 0$) in NdMnAsO. This accounts for why the //ab-O1 spin arrangement is more stable than the //ab-O2 spin arrangement in CeMnAsO, while the opposite is the case for NdMnAsO (**Table 1**).

**B. Experimental implications**

There are two important implications of our work that can be tested experimentally. One deals with the local nature of the DM and BQ exchanges. The antiferromagnetic coupling in the Mn lattice of LaMnAsO is strong since the Mn moments order well above room temperature.[17] Thus, when the moments of a few $Mn^{2+}$ ions are forced to change their directions, the moments of the remaining $Mn^{2+}$ ions will follow them in order to maintain a collinear antiferromagnetic arrangement. For the solid solution $La_{1-x}Ce_xMnAsO$, in which the magnetic ions $Ce^{3+}$ are diluted with nonmagnetic ions $La^{3+}$, the SR is expected to occur as long as the DM exchange of the remaining $Ce^{3+}$ ions with their surrounding $Mn^{2+}$ ions is stronger than the easy-axis anisotropy of the $Mn^{2+}$ moment (i.e., 4.88 x > 0.2). This prediction has recently been confirmed by Tsukamoto *et al.*,[21] who showed that the SR in $La_{1-x}Ce_xMnAsO$ persists for x down to 0.10. Similarly, the solid solution $La_{1-x}Nd_xMnAsO$ is predicted to exhibit the SR for x > 0.01. In addition, the solid solution $Ce_{1-x}Nd_xMnAsO$ is predicted to adopt the //ab-C arrangement for x > 0.22 according to the $E_{DM}$ and $E_{BQ}$ values of CeMnAsO and NdMnAsO in **Table 1**. The other implication is concerned with how to tilt the balance between the DM and BQ exchanges. For the DM exchange to dominate over the BQ exchange, it would be necessary to reduce the $(\vec{S}_0 \cdot \vec{S}_1)^2$ value, and hence the combination of the S=1/2 ion $Ce^{3+}$ with TM ions with S < 5/2 (e.g., S=2 ion such as $Cr^{2+}$ and $Fe^{2+}$) might be interesting.

## 5. Concluding remarks

In summary, our study shows that the SR phenomena of the square lattice antiferromagnets RMnAsO (R = Ce, Nd) are well explained in terms of the DM and BQ exchanges between $R^{3+}$ and $Mn^{2+}$ ions determined in terms of DFT calculations. The symmetry of the DM vector in such systems follows that of the partially filled 4f states of the $R^{3+}$ ions (Eq. 1), and the consideration of the DM and BQ exchanges between RE and TM ions is essential in describing magnetic solids containing both types of ions. Below their SR temperature, the $Mn^{2+}$ and $Ce^{3+}$ moments are orthogonal in CeMnAsO but are collinear in NdMnAsO because the DM interaction dominates over the BQ interaction in CeMnAsO while the opposite is the case for

NdMnAsO. The implications of our findings discussed above need to be tested by further experiments and theoretical analyses.

NdMnAsO. The implications of our findings discussed above need to be tested by further experiments and theoretical analyses.

**Acknowledgments**

The work at North Carolina State University was supported by the Office of Basic Energy Sciences, Division of Materials Sciences, U. S. Department of Energy, under Grant DE-FG02-86ER45259, and also by the computing resources of the NERSC center and the HPC center of NCSU. Work at UVa was supported by the U.S. Department of Energy, Office of Basic Energy Sciences, Division of Materials Sciences and Engineering under Award DE- FG02-10ER46384.

* E-mail: mike_whangbo@ncsu.edu

**References**

(1) Erdös, P., *J. Phys. Chem. Solids*, **1966**, *27*, 1705.

(2) Moriya, T., *Phys. Rev*. **1960**, *120*, 91 (1960).

(3) (a) Shekhtman, L.; Entin-Wohlman, O.; Aharony, A., *Phys. Rev. Lett*. **1992**, *69*, 836. (b) Yildirim, T.; Harris, A. B.; Aharony, A.; Entin-Wohlman, O., *Phys. Rev. B* **1992**, *52*, 10239.

(4) Sólyom, J., *Phys. Rev. B* **1987**, *36*, 8642.

(5) Novák, P.; Chaplygin, I.; Seifert, G.; Gemming, S.; Laskowski, R., *Comput. Mater. Sci*. **2008**, 79, 44.

(6) Furrer, A.; Juranyi, F.; Krämer, K. W.; Strässle, Th., *Phys. Rev. B* **2008**, *77*, 174410.

(7)  Dzyaloshinskii, I., *J. Phys. Chem. Solids* **1958**, *4*, 241.

(8)  Katsura, H.; Nagaosa, N.; Balatsky, A. V., *Phys. Rev. Lett*. **2005**, *95*, 057205.

(9)  Sergienko, I. A.; Dagotto, E., *Phys. Rev. B* **2006**, *73*, 094434.

(10) Onose, Y.; Ideue, T.; Katsura, H.; Shiomi, Y., Nagaosa, N.; Tokura, Y., *Science* **2010**, *329*, 297.

(11) Ginsberg, A. P., *Inorg. Chim. Acta Rev*. **1971**, *5*, 45.

(12) Bencini, A.; Gatteschi, D., *EPR of Exchange Coupled Systems*; Springer, 1989.

(13) Benelli, C.; Gateschi, D., *Chem. Rev*. **2002**, *102*, 2369.

(14) Skanthakumar, S.; Zhang, H.; Clinton, T. W.; Li, W.-H.; Lynn, J. W.; Fisk, Z.; Cheong, S.-W., *Physica C*. **1989**, *160*, 124.

(15) Sachidanandam, R.; Yildirim, T.; Harris, A. B.; Aharony, A.; Entin-Wohlman, O., *Phys. Rev. B* **1997**, *56*, 260.

(16) Marcinkova, A.; Hansen, T. C.; Curfs, C.; Margadonna, S.; Bos, J.-W. G., *Phys. Rev. B* **2010**, *82*, 174438.

(17) Emery, N.; Wildman, E. J.; Skakle, J. M. S.; Mclaughlin, A. C.; Smith, R. I.; Fitch, A. N., *Phys. Rev. B* **2011**, *83*, 144429.

(18) Kimber, S. A. J.; Hill, A. H.; Zhang, Y.-Z.; Jeschke, H. O.; Valentí, R.; Ritter, C.; Schellenberg, I.; Hermes, W.; Pöttgen, R.; Argyriou, D. N., *Phys. Rev. B* **2010**, *82*, 100412.

(19) Kamihara, Y.; Watanabe, T.; Hirano, M.; Hosono, H., *J. Am. Chem. Soc*. **2008**, *130*, 3296.

(20) Ishida, K.; Nakai, Y.; Hosono, H., *J. Phys. Soc. Jpn*. **2009**, *78*, 062001.

(21) Tsukamoto, Y.; Okamoto, Y.; Matsuhira, K.; Whangbo, M.-H.; Hiroi, Z., *J. Phys. Soc. Jpn*. **2011**, *80*, 094708.

(22) Lee, J.-S.; Ji, S.; Lee, S.-H., private communication.

(23) Dai, D.; Xiang, H.-J.; Whangbo, M.-H., *J. Comput. Chem*. **2008**, *29*, 2187.

(24) Buschow, K. H. J., editor, *Concise Encyclopedia of Magnetic and Superconducting Materials, Second Edition*; Elsevier: Armsterdam, 2006, p 150.

(25) (a) Kresse, G.; Hafner, J., *Phys. Rev. B* **1993**, *62*, 558 (1993). (b) Kresse, G.; Furthmuller, J., *Comput. Mater. Sci*. **1996**, *6*, 15. (c) Kresse, G.; Furthmüller, J., *Phys. Rev. B* **1996**, *54*, 11169.

(26) Perdew, J. P.; Burke, K.; Ernzerhof, M., *Phys. Rev. Lett*. **1996**, *77*, 3865.

(27) Dudarev, S. L.; Botton, G. A.; Savrasov, S. Y.; Humphreys, C. J.; Sutton, A. P.; *Phys. Rev. B* **1998**, *57*, 1505.

(28) Ji, S.; Kan, E. J.; Whangbo, M.-H.; Kim, J.-H.; Qiu, Y.; Matsuda, M.; Yoshida, H.; Hiroi, Z.; Green, M. A.; Ziman, T.; Lee, S.-H., *Phys. Rev. B* **2010**, *81*, 094421.

(29) Ben Yahia, H.; Gaudin, E.; Boulahya, K.; Darriet, J.; Son, W.-J.; Whangbo, M.-H., *Inorg. Chem*. **2010**, *49*, 8578.

(30) Blaha, P.; Schwarz, K.; Madsen, G. K. H.; Kvasnicka, D.; Luitz, J., WIEN2k; Vienna University of Technology, 2001 (ISBN 3-9501031-1-2).

(31) The f-electrons of the RE ions are localized than the d-electrons of the TM ions, so the use of a larger U for the RE ion is justified.

(32) Kuneš, K.; Novák, P.; Diviš, M.; Oppeneer, P. M., *Phys. Rev. B* **2001**, *63*, 205111.

(33) Ji, S.; Kan, E. J.; Whangbo, M.-H.; Kim, J.-H.; Qiu, Y.; Matsuda, M.; Yoshida, H.; Hiroi, Z.; Green, M. A.; Ziman, T.; Lee, S.-H., *Phys. Rev. B* **2010**, *81*, 094421.

(34) Xiang, H. J.; Wei, S.-H.; Whangbo, M.-H.; Da Silva, J. L. F., *Phys. Rev. Lett*. **2008**, *101*, 037209.

(35) Xiang, H. J.; Wei, S.-H.; Whangbo, M.-H., *Phys. Rev. Lett*. **2008**, *100*, 167207.

19(36) Koo, H.-J.; Xiang, H. J.; Lee, C.; Whangbo, M.-H., *Inorg. Chem.* **2009**, *48*, 9051.

(37) (a) Dai, D.; Whangbo, M.-H., *J. Chem. Phys.* **2001**, *114*, 2887. (b) Dai, D.; Whangbo, M.-H., *J. Chem. Phys.* **2003**, *118*, 29.

(38) Whangbo, M.-H.; Koo, H.-J.; Dai, D., *J. Solid State Chem.* **2003**, *176*, 417.

(39) Xiang, H. J.; Lee, C.; Whangbo, M.-H. *Phys. Rev. B*: *Rapid Commun.* **2007**, *76*, 220411(R).

(40) Koo, H.-J.; Whangbo, M.-H. *Inorg. Chem.* **2008**, *47*, 128.

(41) Koo, H.-J.; Whangbo, M.-H. *Inorg. Chem.* **2008**, *47*, 4779.

(42) Koo, H.-J.; Lee, C.; Whangbo, M.-H.; McIntyre, G. J.; Kremer, R. K., *Inorg. Chem.* **2011**, *50*, 3582.

Table 1.   The relative energies E (in meV per FU) of the //ab-C, //ab-O1 and //ab-O2 spin arrangements of RMnAsO (R = Ce, Nd) obtained from DFT+U+SOC calculations and the associated DM and BQ exchange energies (in meV per FU) between the $R^{3+}$ and $Mn^{2+}$ ions.

|              | CeMnAsO | NdMnAsO |
|--------------|---------|---------|
| E(//ab-C)    | 0       | 0       |
| E(//ab-O1)   | -4.81   | +18.39  |
| E(//ab-O2)   | +4.96   | +17.70  |
| $E_{DM}$     | -4.88   | -0.35   |
| $E_{BQ}$     | -0.07   | -18.05  |

Table 2.    The distances between the magnetic ions in the Heisenberg exchange paths $J_1 - J_5$ of CeMnAsO and the values of $J_1 - J_5$ obtained from the mapping analysis based on the DFT+U calculations.

| Exchange paths | Distance (Å) | $J_i$ (meV) |
|---|---|---|
| $J_1$ (within each Mn layer) | Mn-Mn = 2.884 | +13.1 |
| $J_2$ (within each Mn layer) | Mn-Mn = 4.078 | +5.9 |
| $J_3$ (between adjacent Ce layers) | Ce-Ce = 3.722 | +92.8 |
| $J_4$ (within each Ce layer) | Ce-Ce = 4.078 | −120.3 |
| $J_5$ (between adjacent Ce and Mn layers) | Ce-Mn = 3.871 | −11.2 |

**Figure captions**

Figure 1.  (a) The arrangement of the $Nd^{3+}$ ions above and below the square lattice of $Cu^{2+}$ ions in $Nd_2CuO_4$, where the $Nd^{3+}$ and $Cu^{2+}$ ions are represented by white and pink spheres, respectively. The two Nd atoms located above and below each $Cu_4$ square forms a $Cu_4Nd_2$ square bipyramid. (b) The orthogonal arrangement between the $Nd^{3+}$ and $Cu^{2+}$ moments in $Nd_2CuO_4$, where the larger and smaller circles represent the $Nd^{3+}$ and $Cu^{2+}$ ions, respectively.

Figure 2.  (a) A perspective view of the crystal structure of RMnAsO (R = Ce, Pr, Nd), in which each square sheet of high-spin $Mn^{2+}$ ($d^5$, S = 5/2) ions parallel to the ab-plane is sandwiched between two square sheets of $As^{3-}$ ions to form an As-Mn-As slab (Mn = red circle, As = purple circle). Similarly, each square sheet of $O^{2-}$ ions parallel to the ab-plane is sandwiched between two square sheets of $R^{3+}$ ions to form a R-O-R slab (O = cyan circle, R = green circle). In RMnAsO the As-Mn-As slabs alternate with the R-O-R slabs along the c-direction such that each R is located above the center of an $As_4$ square, and each As above the center of a $R_4$ square. (b) A perspective view of the $R^{3+}$ and $Mn^{2+}$ ion arrangement in RMnAsO. The $R^{3+}$ ions above and below the square sheet of the $Mn^{2+}$ ions occupy the positions of neighboring $Mn_4$ squares, each R atom forming $Mn_4R$ square pyramid.

Figure 3.  (a) The arrangement of the $R^{3+}$ ions above and below the square lattice of $Mn^{2+}$ ions in RMnAsO (R = Ce, Nd), where the $R^{3+}$ and $Mn^{2+}$ ions are represented by white and cyan spheres, respectively. Each R atom located above or below every $Mn_4$ square forms a $Mn_4R$ square pyramid. (b – d) The orthogonal and collinear arrangements between the $R^{3+}$ and $Mn^{2+}$ moments in RMnAsO (R = Ce, Nd), where the larger and smaller circles represent the $R^{3+}$ and $Mn^{2+}$ ions, respectively. The //ab-C, //ab-O1 and //ab-O2 arrangements of the $R^{3+}$ and $Mn^{2+}$ moments are presented in (b), (c) and (d), respectively. In the square lattice of the $Mn^{2+}$ ions, the $Mn^{2+}$ moments are antiferromagnetically coupled. The $R^{3+}$ and $Mn^{2+}$ moments lie in the ab-plane (i.e., the plane of the square lattice). The $R^{3+}$ moments located above the square lattice of the $Mn^{2+}$ ions are antiparallel to those located below the square lattice of the $Mn^{2+}$ ions.

Figure 4. The total DOS and the PDOS plots of CeMnAsO obtained from the DFT+U+SOC calculations for the //ab-C spin arrangement. The bottom of the empty bands is taken as the Fermi level.

Figure 5. The total DOS and the PDOS plots of NdMnAsO obtained from the DFT+U+SOC calculations for the //ab-C spin arrangement. The bottom of the empty bands is taken as the Fermi level.

Figure 6. The five spin-exchange paths $J_1 - J_5$ of RMnAsO. For simplicity, only the $Mn^{2+}$ and $R^{3+}$ ions are shown as small and large circles. The numbers 1 – 5 refer to the spin-exchange paths $J_1 - J_5$, respectively. The $J_5$ path is the Heisenberg exchange $J_{0i}$ between the $R^{3+}$ and $Mn^{2+}$ ions.

Figure 7. The six ordered spin states of RMnAsO (R = Ce, Nd), where the shaded circles represent the up-spin R and Mn sites, and the unshaded circles the down-spin R and Mn sites. The numbers in each parenthesis refer to the relative energies (in meV per FU) determined for CeMnAsO from the DFT+U calculations.

Figure 8. PDOS plots for (a) the Ce 4f states of CeMnAsO and (b) the Nd 4f states of NdMnAsO obtained from the DFT+U+SOC calculations for the //ab-C spin arrangement, where the numbers 0, ±1, ±2 and ±3 refer to the $l_z$ values of the $Ce^{3+}$ and $Nd^{3+}$ ions. The top of the filled bands is taken as the Fermi level. In each diagram, the upper and lower panels represent the up-spin and down-spin states, respectively. In each panel, the vertical PDOS axis covers from 0 to 1.5 states/eV/atom, and the numbers on the horizontal energy axis are in eV.

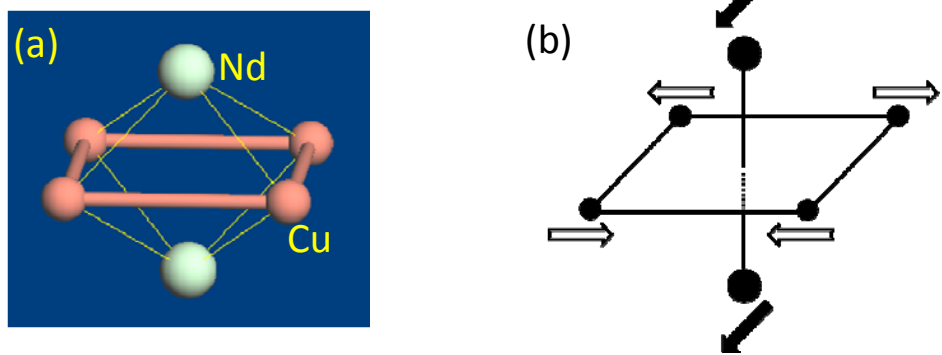

Figure 1.

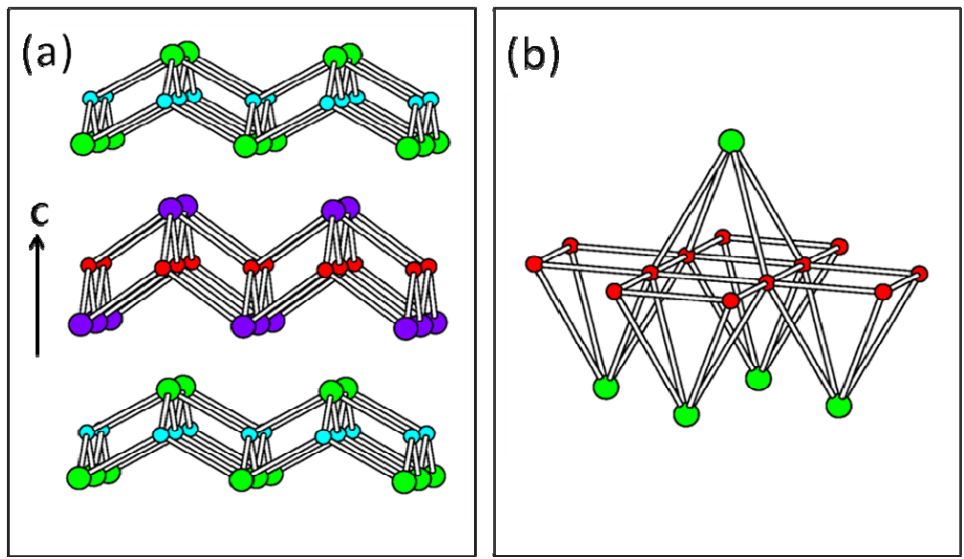

Figure 2

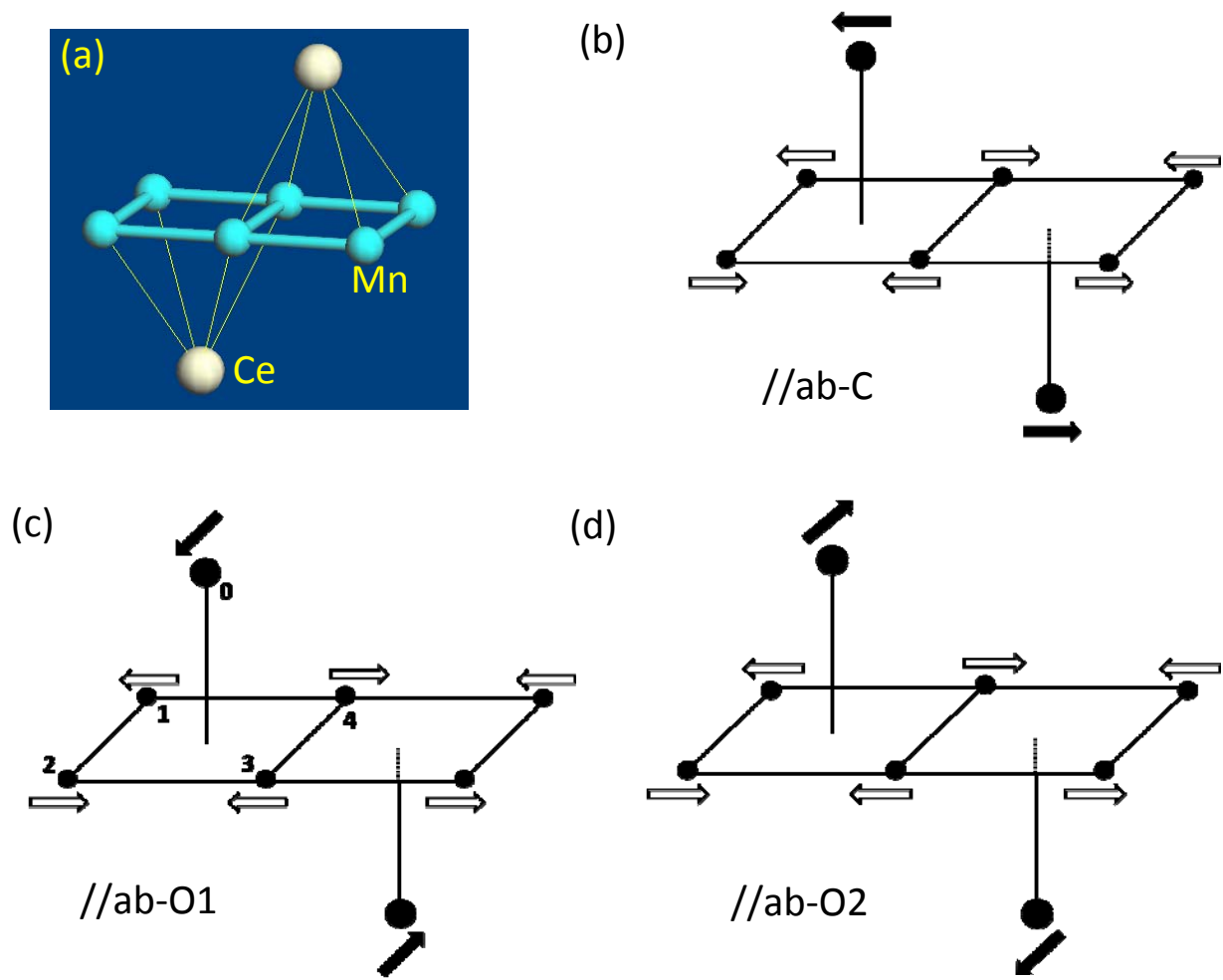

Figure 3.

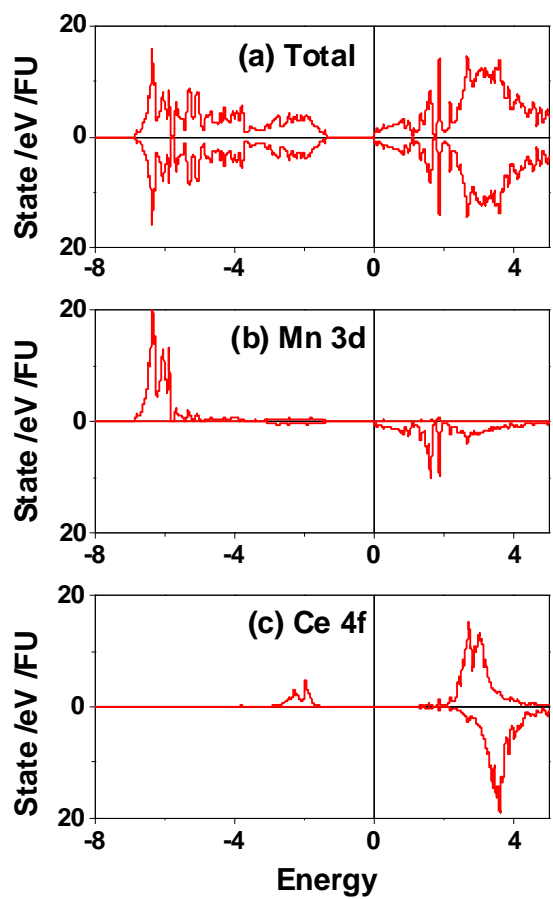

Figure 4.

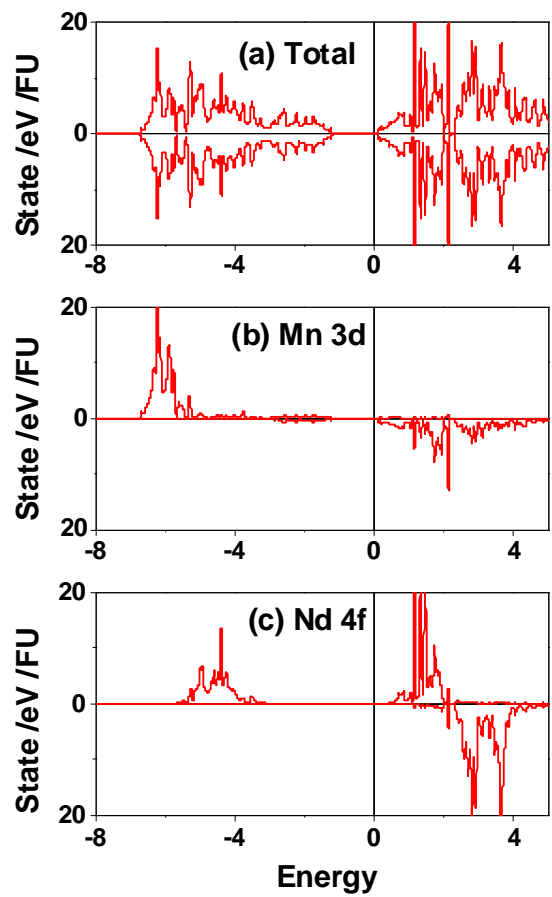

Figure 5.

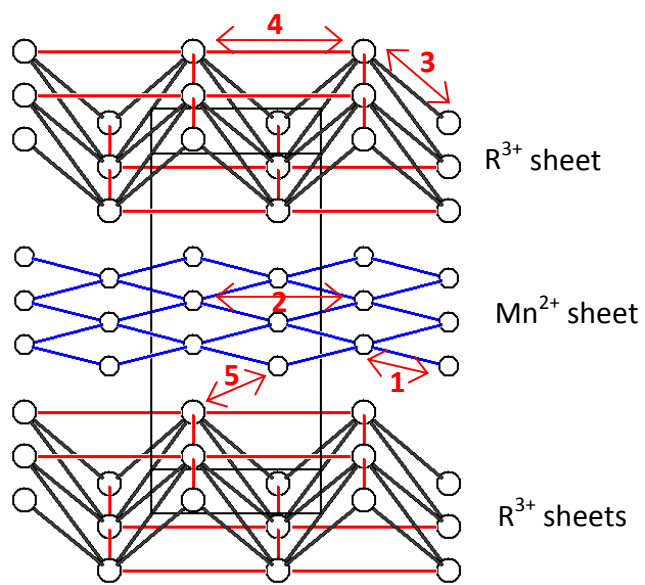

Figure 6.

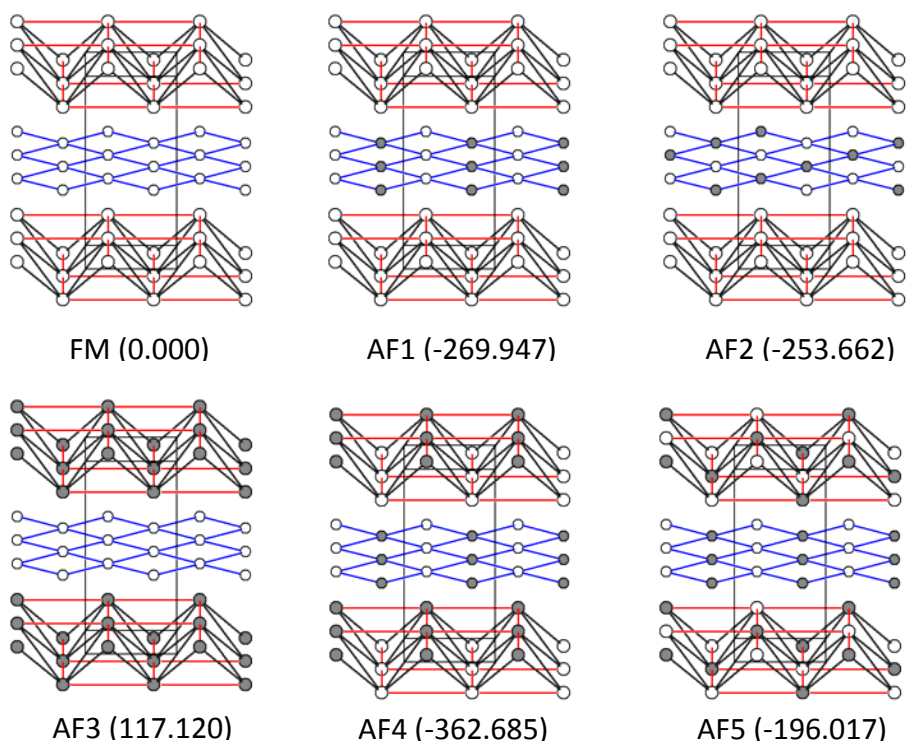

FM (0.000)  AF1 (-269.947)  AF2 (-253.662)

AF3 (117.120)  AF4 (-362.685)  AF5 (-196.017)

Figure 7.

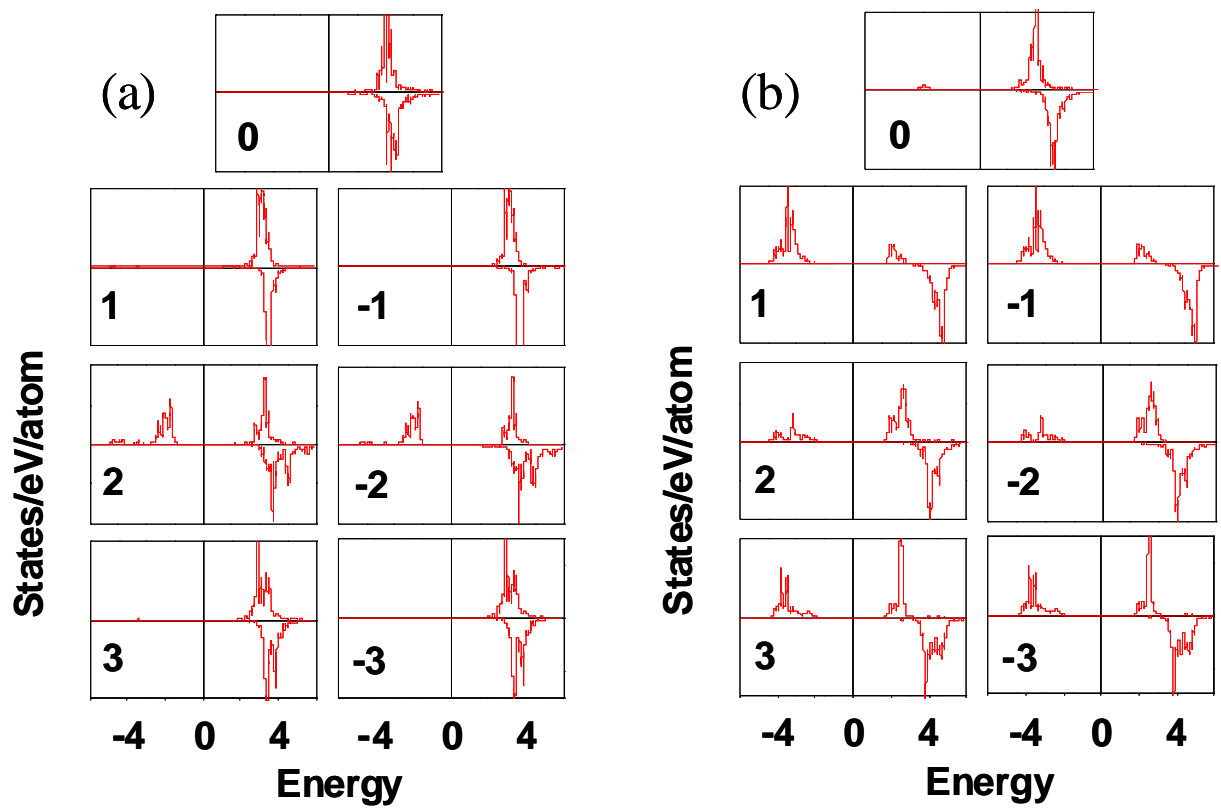

Figure 8.

**Synopsis**

The spin reorientation phenomenon of the square-lattice antiferromagnets RMnAsO (R = Ce, Nd) was investigated by calculating the Dzyaloshinskii-Moriya and biquadratic exchange interactions between the rare-earth and transition-metal ions ($R^{3+}$ and $Mn^{2+}$, respectively) and by analyzing the symmetry of the Dzyaloshinskii-Moriya exchange. The observed spin reorientation is controlled by the Dzyaloshinskii-Moriya and biquadratic exchanges, and the symmetry and strength of the Dzyaloshinskii-Moriya exchange are determined mainly by the partially filled 4f states of the $R^{3+}$ ions.

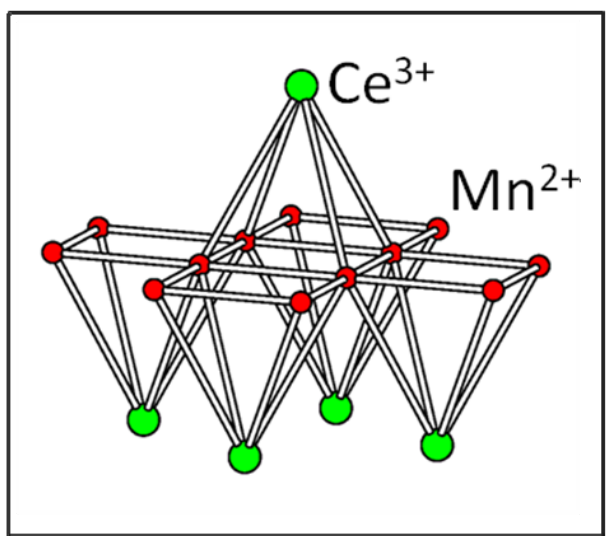
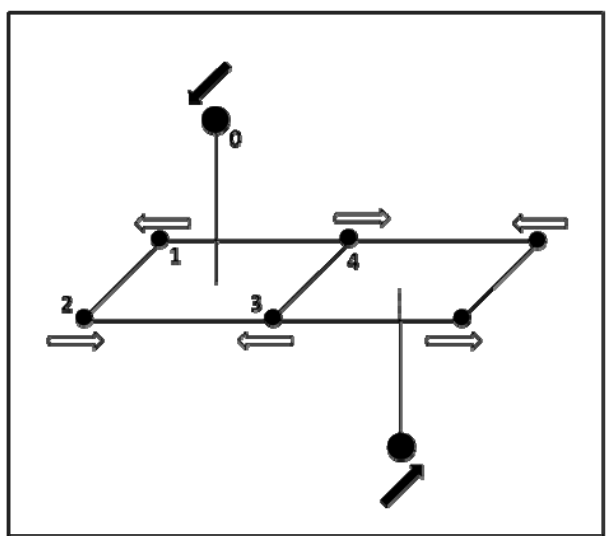